# Quantum Diplomacy within the Southeast Asia Quantum Ecosystem

Pak Shen Choong[a*] (0000-0002-8271-0430), Nurisya Mohd Shah[a,b] (0000-0001-7783-1118), Yung Szen Yap[c] (0000-0003-0201-4038)

[a] Institute for Mathematical Research, Universiti Putra Malaysia, 43400 UPM Serdang, Selangor Darul Ehsan.

[b] Department of Physics, Faculty of Science, Universiti Putra Malaysia, 43400 UPM Serdang, Selangor Darul Ehsan.

[c] Department of Physics, Faculty of Science, Universiti Teknologi Malaysia, 81310 UTM Johor Bahru, Johor Darul Ta'zim.

[*] Corresponding author: pakshen@upm.edu.my

**Abstract**

Amid the International Year of Quantum Science and Technology 2025 (IYQ 2025), a significant portion of global funding has been dedicated to various quantum initiatives, with over 30 countries announcing their respective quantum strategies. Within the Southeast Asia context, Singapore, Thailand, and the Philippines have launched their respective quantum strategies and roadmaps. Meanwhile, six out of eleven Southeast Asia countries have expressed interest in formulating a regional quantum ecosystem to pursue a set of common goals. Quantum technologies, though still in their infancy within the second quantum revolution, have advanced rapidly in recent years. Due to their dual-use nature, quantum technologies are considered emerging and disruptive, often raising concerns from the cybersecurity perspective. While several discussions regarding Malaysia's quantum initiative and strategy are ongoing, it is vital to broaden the conversation and position Malaysia within the regional ecosystem. This paper provides an overview of Malaysia's quantum landscape and a summary of the regional initiatives since the establishment of Southeast Asia Quantum Network. We then analyse Malaysia's strengths in quantum research and provide four recommendations to strengthen the regional ecosystem.

# 1.0 Introduction

## 1.1 From science diplomacy to tech diplomacy

According to Article 27 from the Universal Declaration of Humans Rights (1948), one of the basic human rights is the freedom to participate in science and enjoy its benefits. This becomes the underlying principle of *science diplomacy*, where science is seen as the soft power to facilitate diplomatic conversations in such a way that the political climates between nations will not halt the development of science and technologies. While the significance of science diplomacy can be traced back to as early as the pre-20th century in cartography, its contributions can be broadly broken down into three categories:

1. *Standardization*, for instance the International System of Units (International Bureau of Weights and Measures, 2022);
2. *Conflict resolution and crisis management*, for instance the disarmament of nuclear weapons after World War II through the Atoms for Peace initiative (Eisenhower, 1953), and the coordinated global pandemic management by the World Health Organization (WHO) (Vargha & Wilkins, 2023);
3. *Scientific collaboration*, for instance the large-scale atomic and high energy physics experiments by the European Organization for Nuclear Research (CERN).

The interactions between science and diplomacy are often traced back to a clear definition provided by the joint work of the Royal Society and American Association for the Advancement of Science (AAAS), where three dimensions of science diplomacy are defined, namely (The Royal Society & AAAS, 2010)

1. *Science in diplomacy*, which focuses on scientific contributions in policy making;
2. *Science for diplomacy*, which focuses on using science as the soft power and currency in diplomatic negotiations and hard power situations;
3. *Diplomacy for science*, which focuses on utilizing diplomatic and political ties to facilitate international scientific collaborations.

While science diplomacy highlights the importance of science in diplomatic situations due to its universality, it is important not to over-politicize science in a political scenario.

In recent years, the world has become more and more fragmented due to various geopolitical and geoeconomic tensions. Additionally, the rapid development of emerging and disruptive technologies (NATO, 2025) with dual use potentials, such as artificial intelligence and quantum technologies, has intensified the tension between the spirit of open science and national security concerns. In this fast-paced era, science diplomacy has further evolved into three different approaches. The first approach further finetunes the definition of science diplomacy through a fourth dimension, namely *diplomacy in science*, i.e. to use diplomatic skills in science (Gjedssø Bertelsen et al., 2025). The second approach looks at the broader aspects of science diplomacy and reframes it into how science and diplomacy impact one another (AAAS & The Royal Society, 2025). The third approach adds anticipation into science diplomacy to encourage forward thinking and risk mitigation into diplomatic actions (GESDA, 2025).

Within the broader spectrum of science diplomacy, *tech diplomacy* also started to gain traction through the TechPlomacy initiative (Klynge, Ekman, & Waedegaard, 2020) by a Danish diplomat Casper Klynge. Recognizing the growing power of non-state actors, tech diplomacy attempts to address the complexities and challenges of multilateralism coming from a new dimension of power dynamics, called the *innovative power*. Tech diplomacy scrutinizes not only the technology itself, but also tries to identify actors and their interactions, and how these actors are ordered due to the background and worldview they are adopting (Rushton & Williams, 2012). Collectively known as *the analytical triangle*, i.e. *technology*, *agency*, and *order*, tech diplomacy aims to redefine power dynamics across state and non-state actors, address the inequalities that come with the technological disruption, and establish a regulatory and governance framework to mitigate potential risks (Bjola & Kornprobst, 2025).

**1.2 Quantum diplomacy**

As pointed out by Dowling and Milburn (2003), the second quantum revolution is a result of the current technologies to control quantum systems and harness various quantum effects, such as quantum superposition and entanglement to our benefits. Current quantum technologies can be broadly categorized into three applications, i.e. *quantum communication*, *quantum sensing*, and *quantum computing*. These technologies progress at different rates and maturity levels, which can be measured by the Technology Readiness Levels (TRLs) (OECD, 2025). They possess superiorities over the classical technologies in certain tasks, collectively known as the *quantum advantage*, which usually manifests itself either in the form of computational speedup, entanglement-enhanced sensing sensitivity, or cryptographic security. At the current stage, the identification of the areas where quantum technologies may excel requires empirical evidence, mathematical analysis, or benchmarking of resources, although these classical analyses alone may be insufficient to completely capture the potentials of quantum technologies (Huang et al., 2025).

Despite most quantum technologies have low TRLs and only getting validated either in laboratory or relevant environment (OECD, 2025), quantum technologies have already demonstrated immense dual-use potentials in *quantum warfare* (Krelina, 2021; Krelina, 2025). One of the most prominent examples is the breaking of Rivest-Shamir-Adleman (RSA) encryption scheme due to Shor's algorithm (Shor, 1994; Shor, 1997), even though quantum warfare may extend beyond the cybersecurity domain into military navigation and surveillance due to quantum sensing, quantum imaging, and quantum radar. Because of this, multiple quantum-developed nations, such as the United Kingdom[1], United States[2], and the European Union[3] have listed quantum technologies under the export control of dual-use items. While being seen as a necessary move to safeguard quantum technological sovereignty and mitigate possible threats, this export control further escalates the *quantum divide* in science, technologies, between countries, and within the societies (Ten Holter et al., 2022; Gercek & Seskir, 2025).

---

[1] NTE 2024/04: The Export Control (Amendment) Regulations 2024.
[2] Bureau of Industry & Security, U.S. Department of Commerce.
[3] Directorate-General for Trade & Economic Security. 2025 update of the EU control list of dual-use items.

*Quantum diplomacy*, a terminology first coined by the former U.S. Secretary of State George P. Shultz in his conversation with a theoretical physicist friend Sidney Drell (Shultz, 1997), analogized the disruptions of the information age on diplomacy to the physical process of quantum measurements on quantum systems. The definition of quantum diplomacy as a form of tech diplomacy was later given by Randolph Mank in 2017. From his definition, the foreign policy surrounding quantum technologies should focus on securities (particularly risks coming from non-state actors) and diplomatic opportunities from the convergence of quantum technologies and artificial intelligence. Since 2021, Geneva Science Diplomacy and Anticipator (GESDA) anticipated quantum technology as one of the impactful breakthroughs in the future. As a result of several multilateral conversations and collaborations, the Open Quantum Institute (OQI) was incubated at GESDA and later piloted at CERN as an anticipation instrument with four objectives, focusing on navigating the quantum divide through multilateral governance for the sustainable development goals (SDGs) (GESDA & OQI, 2023; OQI, 2024). The first Quantum Diplomacy Symposium further highlighted that security, supply chain and sovereignty, access and education, human capital, and human agency as the five challenges for the future multilateral governance of quantum computing (GESDA, 2024).

Learning from the disruptions that come with the technological diffusion of artificial intelligence into the society (Feng, Cramer, & Mans, 2022), the community from Ethical, Legal, and Social Aspects (ELSA) of quantum technologies studies how to navigate the quantum divide and mitigate unintended risks from quantum technologies through several approaches, particularly Responsible Research and Innovation (RRI) principles, Diversity, Equity, and Inclusion (DEI) considerations, and lastly workforce development through outreach efforts aiming at underserved regions (Schmidt et al., 2025). The ELSA of quantum technologies are further guided by the Safeguarding, Engaging, and Advancing (SEA) framework, ensuring that quantum technologies are being *advanced* when risks are *safeguarded*, while stakeholders are *engaged* in the innovation process (Kop et al., 2023; Gasser, De Jong, & Kop, 2024).

### 1.3 Motivations of this paper

In short, quantum diplomacy respects the human right to participate in quantum science and enjoy the technological benefits. However, the growing innovative power from the non-state actors, coupled with the dual-use potential of quantum technologies, have created a fragmented interest in quantum technologies from different stakeholders. The goal of quantum diplomacy is to reconvene various stakeholders from the quantum ecosystem, either locally, regionally or globally, to soften the hard power situations due to geopolitical tension, national security and export control, technological sovereignty, and the imbalanced budget for quantum science and technology[4]. Such multilateral dialogues redirect the attention to develop quantum science and technology for humanity, hence narrowing the quantum divide and eventually addressing global challenges through partnerships with a set of common goals, i.e. the SDGs.

Within the topic of navigating quantum divide, Gercek and Seskir (2025) discussed the national adoption and development of quantum technologies through path dependency theory (Liebowitz & Margolis, 1995). This is relevant to the Southeast Asia countries, since most countries[5] cannot afford to diversify their limited budget to develop quantum computing, quantum communication, and quantum sensing in parallel. As a matter of fact, if we zoom out to the Global South setting, other pressing socio-economical issues often overwhelm the government budget, leaving little to none of the funding to develop any quantum initiative. Presented as one of the viable strategies, establishing a regional quantum ecosystem with a set of mutual benefits will alleviate the financial strain, reinforce each other's path dependency based on their respective strengths in quantum technological development, and make up for the weaknesses that comes with the prioritization of specific development pathways. This strategy requires coordination among the participating countries and stakeholders, further illustrates the importance of quantum diplomacy in facilitating the regional development of quantum science and technologies.

---

[4] QURECA, Quantum Initiatives Worldwide 2025: https://www.qureca.com/quantum-initiatives-worldwide/

The purpose of this paper is two-fold. First, we aim to break down Malaysia's current quantum landscape to better position Malaysia within the Southeast Asia quantum ecosystem. Second, we provide four recommendations to strengthen the Southeast Asia quantum ecosystem from the viewpoint of a quantum scientist. These recommendations are derived to mitigate quantum hype and misinformation, devise standardization and benchmark metric, improve the quality of quantum research, provide educational access to everyone, sustain the grassroot operations, and finally assist market penetration in the region. In Section 2, we explore Malaysia's quantum landscape in terms of academia, government, and industry. Section 3 summarizes the philanthropic grassroot initiatives during the International Year of Quantum Science and Technology (IYQ 2025)[6], highlighting the possibility to develop a Southeast Asia quantum ecosystem. In Section 4, we provide some recommendations to further support the development of this regional ecosystem.

**2.0 Malaysia's quantum landscape**

**2.1 Academia**

For the past two decades, the groundwork of quantum technologies in Malaysia was laid in the free space quantum key distribution demonstrations (Abdul Khir et al., 2013) by MIMOS, a public company that serves as Malaysia's national applied research and development centre. From 2011 to 2014, a five-node quantum cryptography network was maintained by MIMOS until project completion. Within the project's duration, MIMOS filed 22 patents and was considered as the top 6$^{th}$ patent applicant worldwide (Lewis & Travagnin, 2018). At the same time, a series of efforts in quantum information and quantum foundations were launched by the theoretical and computational physics research group in Universiti Putra Malaysia (UPM), notably through the Expository Quantum Lecture Series (EQuaLS), whereby the first event was held in 2007[7]. In 2023, the Quantum Information Meetup at Xiamen University Malaysia (MyQI & IFM, 2023) catalysed the formation of Malaysia's quantum grassroot initiative, called Malaysia Quantum Information Initiative (MyQI). As a coalition of quantum scientists in Malaysia, MyQI has been active in advising various governmental

---
[5] Singapore developed a quantum strategy since 2007.
[6] https://quantum2025.org/
[7] https://einspem.upm.edu.my/equals1/

ministries and advancing the quantum development roadmap in Malaysia. In addition to the grassroot initiative, individual universities are establishing research labs and centers of excellence for quantum theories and technologies, such as

1. IIUM Photonics Quantum Center (iPQC);
2. UM Center of Excellence Quantum Information Science and Technology (CoE QIST);
3. UniMAP Centre of Excellence for Advanced Communication Engineering (CoE ACE);
4. UPM Pusat Teknologi dan Pengurusan Kriptologi Malaysia (Malaysia Cryptology Technology and Management Centre).

Figure 1 shows the distribution of quantum scientists across Malaysia, while Figure 2 presents the estimated percentage of Malaysian quantum scientists by their research areas. Table 1 provides another perspective based on the quantum research areas and participating universities.

At the moment, quantum physics and quantum information courses are introduced only under the physics program in most universities. A market survey on a computer science bachelor degree in quantum computing technology is currently conducted by Universiti Teknikal Malaysia Melaka (UTeM).

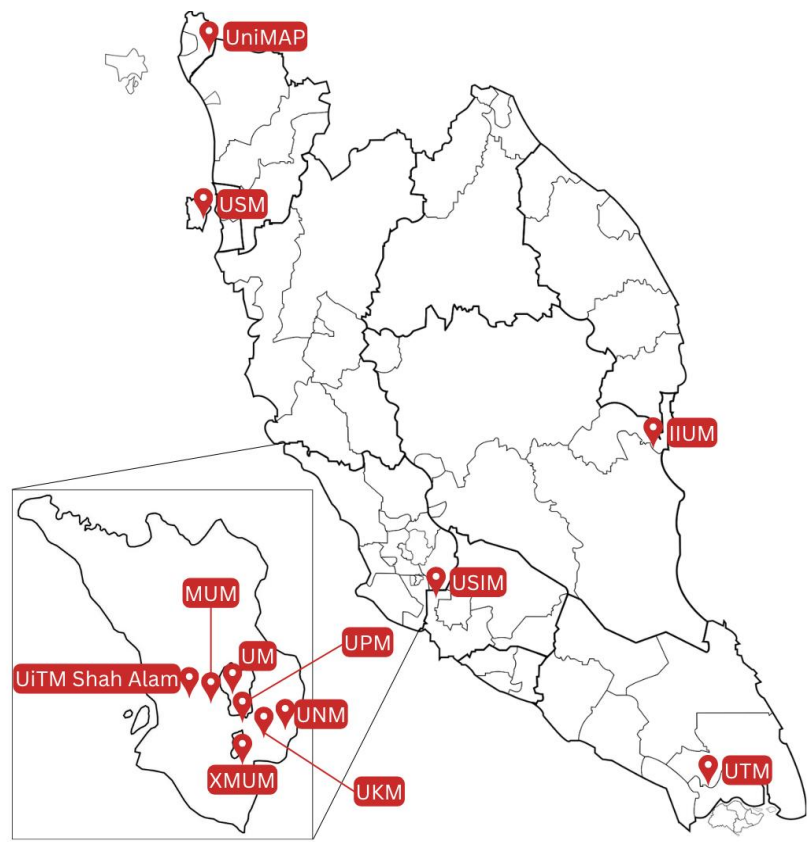

Figure 1: The distribution of quantum scientists in Malaysia according to their universities. To our best knowledge, there is no quantum scientist in East Malaysia.

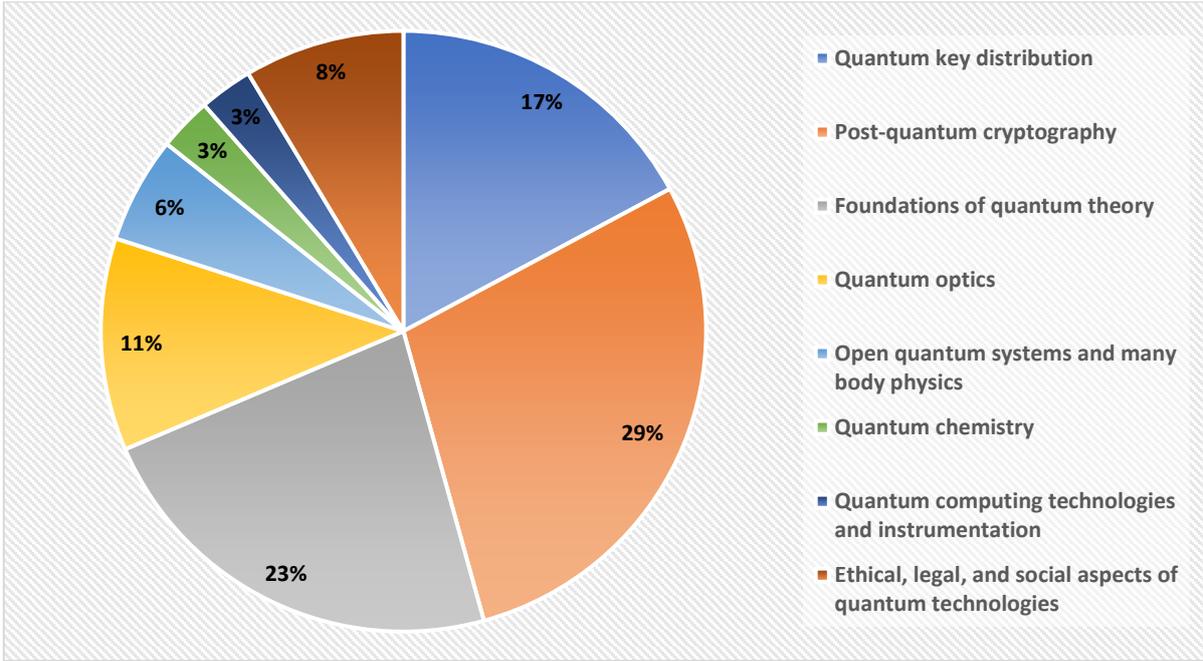

Figure 2: Rough estimates on percentages of researchers working in the designated research areas in Malaysia.

Table 1: Quantum research areas and participating universities.

| Quantum research areas | Participating universities |
|---|---|
| Quantum communication and quantum-safe network | 1. Universiti Sains Islam Malaysia (USIM)<br>2. International Islamic University Malaysia (IIUM)<br>3. Universiti Malaysia Perlis (UniMAP)<br>4. Universiti Putra Malaysia (UPM) |
| Quantum theory and algorithms | 1. Universiti Putra Malaysia (UPM)<br>2. Xiamen University Malaysia (XMUM)<br>3. Universiti Teknologi MARA, Shah Alam (UiTM)<br>4. Universiti Sains Islam Malaysia (USIM)<br>5. Universiti Kebangsaan Malaysia (UKM)<br>6. Universiti Malaya (UM)<br>7. International Islamic University Malaysia (IIUM)<br>8. Universiti Sains Malaysia (USM)<br>9. University of Nottingham Malaysia (UNM) |
| Quantum computing technologies and instrumentation | 1. Universiti Teknologi Malaysia (UTM) |
| Ethical, legal, and social aspects of quantum technologies | 1. Monash University Malaysia (MUM)<br>2. International Islamic University Malaysia (IIUM)<br>3. Universiti Putra Malaysia (UPM) |

**2.2 Policy landscape**

According to the Science & Technology Foresight Malaysia 2050 report (2017) and 10-10 Malaysian Science, Technology, Innovation and Economic (MySTIE) Framework (2020) developed by Academy of Sciences Malaysia (ASM), quantum technologies is recognized as one of the key science and technology drivers under the category of *Advanced Intelligence Systems*. Even though there is no concrete policy on quantum technologies, various ministries such as the Ministry of Science, Technology, and Innovation (MOSTI), Ministry of Digital, National Cyber Security Agency (NACSA), Malaysian Communications and Multimedia Commission (MCMC), and Ministry of Higher Education (MoHE) have expressed their interests to develop a roadmap for quantum technologies and education with MyQI.

In 2024, CyberSecurity Malaysia released a Post-Quantum Cryptography (PQC) Migration Framework[8] to transition Malaysia into a quantum-safe nation, which catalysed the establishment of Pusat Teknologi dan Pengurusan Kriptologi Malaysia (Malaysia Cryptology Technology and Management Centre). On 25 February 2025, MIMOS officially launched the MIMOS Quantum Intelligence Center[9] with the aim to establish a "quantum valley" in collaboration with SDT Inc. of Korea. Meanwhile, National Metrology Institute of Malaysia (NMIM) is interested in quantum metrology and its applications. During the National Digital Economy and 4IR Council Meeting (MED4IRN) on 2 October 2025, the Prime Minister of Malaysia, YAB Dato' Seri Anwar Bin Ibrahim endorsed the proposed National Quantum Policy and the establishment of a National Quantum Task Force[10].

**2.3 Quantum industry, quantum startup and supply chain**

Malaysia has a national startup initiative, called MYStartUp[11] and a sandbox for startups, called National Innovation and Technology Sandbox (NTIS)[12]. So far, there are five quantum startups in Malaysia, i.e. SEA Quantum[13], Lestari Alam Sekitar[14], Smart Tech Asia Pacific[15], Qubios[16], and MolDesignX[17]. Meanwhile, Nano Malaysia[18] is a startup company which is curious about quantum technologies.

From Malaysian Investment Development Authority (MIDA) and Iskandar Regional Development Authority (IRDA), a tax incentive package for supply chain related to quantum computing technologies was announced on 8 January 2025 under the Johor-Singapore Special Economic Zone (JS-SEZ). Some multi-national corporations in Malaysia which have branches or departments in other countries are involved in the supply chain of quantum computing technologies, such as Samtec, Keysight, and Rohde & Schwarz.

---

[8] CD-4-RPT-2724-PQCMigration-V1: https://mykripto.cybersecurity.my/index.php/files/109/Post-Quantum/18/Post-quantum-Cryptography-Migration-Framework.pdf
[9] https://mimosquantum.my/
[10] https://www.freemalaysiatoday.com/category/nation/2025/10/02/govt-to-establish-sovereign-ai-cloud-says-pm
[11] https://www.mystartup.gov.my/home
[12] https://sandbox.gov.my/
[13] https://seaquantum.ai/
[14] https://lestarian.com/
[15] https://smarttechap.com/
[16] https://qubios.io/
[17] https://moldesignx.com/
[18] https://nanomalaysia.com.my/

From the existing industries, Telekom Malaysia is exploring quantum communication systems, while Payment Network Malaysia (PayNet) is studying quantum use cases in finance technologies.

## 3.0 Opportunities for a regional quantum ecosystem

On 25-26 April 2024, a group of quantum scientists from Malaysia, Thailand, Philippines, Indonesia, and Singapore convened to express a common interest in building a regional quantum ecosystem. During the first meetup, the scientists identified four common goals well within the interests of respective nations (Choong & Meevasana, 2025), i.e.

1. Quantum workforce development;
2. Demonstrable quantum use cases relevant to the region;
3. Partnerships between academia, industries, and government;
4. Science communication on quantum technologies.

Resulting from the working group discussion, also known as the *Bangkok Manifesto*, a regional grassroot initiative called Southeast Asia Quantum Network [19] was formed. The ASEAN Quantum Summit 2025 [20] was planned to be hosted in Malaysia from 10-12 December 2025, an event aimed at uniting all stakeholders to initiate a regional quantum ecosystem.

Concurrently, regional collaborative quantum efforts were being amplified in 2025 as well. From 30 July to 1 August 2025, the Quantum Society of the Philippines (QCSP) organized the first quantum conference in the Philippines, called Quantum Information, Science, and Technology Conference in the Philippines 2025 (QISTCon.ph 2025) [21]. The Quantum Diplomacy Game [22], a participatory diplomatic scenario on quantum technologies developed by OQI, was played during the conference. From 3-5 August 2025, a conjoint OQI-supported hackathon and conference, called Southeast Asia Quantum Hackathon 2025 (SEA Quantathon 2025) [23] and Quantum Technology Research Initiative 2025 (QTRi 2025) [24] were organized in Thailand. In both events, industry players and policymakers were invited alongside quantum scientists. Even though the aforementioned quantum efforts were mainly derived from the strong interests of each country to develop its respective national quantum strategy, it is noteworthy that all three regional quantum events were inspired and supported by the Strengthening and Entangling Global Quantum Roots (SEGQuRo) project [25] and OQI. Table 2

---

[19] The grassroot initiative was called ASEAN Quantum Network during the early days and renamed into Southeast Asia Quantum Network for a similar geographical representation of the community: https://worldquantumday.org/news/the-southeast-asia-world-quantum-day-initiatives
[20] https://quantum2025.org/iyq-event/asean-quantum-summit/
[21] https://www.qistcon.ph/
[22] https://open-quantum-institute.cern/quantum-diplomacy-game/
[23] https://qtric.sut.ac.th/quantathon2025/
[24] https://qtric.sut.ac.th/qtri2025/
[25] https://www.itas.kit.edu/english/projects_sesk24_segquro.php

summarizes the current quantum strategy or initiative developed by each Southeast Asia country.

Table 2: Brief summary on regional quantum strategies and initiatives

| Country | Quantum strategies / initiatives |
|---|---|
| Singapore | National Quantum Strategy[26] by the National Quantum Office |
| Malaysia | Malaysia Quantum Information Initiative (MyQI)[27] |
| Thailand | Quantum Technology Research Initiative Collaboration (QTRic)[28] |
| Philippines | Quantum Computing Society of the Philippines (QCSP)[29] |
| Indonesia | Indonesian Quantum Initiative (IQI)[30], National Research and Innovation Agency: Research Center for Quantum Physics (BRIN-Q)[31] |
| Vietnam | VNQuantum[32], Quanova[33] |

**4.0 Recommendations to support the regional quantum ecosystem**

According to AWO (2024), Global South typically follows three different approaches to develop quantum computing, i.e. to build, to procure, or to access by cloud services. The positioning towards which approach greatly depends on the national funding and quantum strategy, available local talents and experts, or the quantum sovereignty and geopolitical alliances one would like to develop. It is possible for a country to switch from one approach to another over short-, mid-, or long-term goals.

Looking at Malaysia's quantum landscape, Malaysia progressed greatly in the migration towards post-quantum cryptography, a strategy suggested by most international organizations such as OECD and OQI to secure the cyberspace. Additionally, Malaysia developed a strong research history in quantum communication and quantum theory. With our current talent pool, we recommend three possible pathways to quickly develop the quantum research in Malaysia, particularly in quantum key distribution technologies, quantum algorithms and use cases, and quantum error correction. In the meantime, Malaysia can focus on quantum computing technologies as a mid- to long-term goal.

When we zoom out to the bigger picture of a Southeast Asia regional quantum ecosystem, it is not difficult to realize that there is one hurdle we need to overcome, i.e. the quantum divides among the Southeast Asian countries, since different countries began their quantum journeys at different times and progressed at different rates with different priorities. Looking at quantum diplomacy from a scientist point of view, we recommend four actions to support the regional quantum ecosystem.

---

[26] https://nqo.sg/nqs/
[27] https://www.myqi.my/
[28] https://qtric.sut.ac.th/
[29] https://www.qcsp.ph/
[30] https://iqi.cat/en/
[31] https://quantumresearch.id/
[32] https://www.vnquantum.org/
[33] https://quanova.org/

**4.1 Engaging stakeholders**

In 2022, World Economic Forum (WEF) released a report on quantum computing governance principles. From the report, the stakeholders of quantum computing are generally categorized into five groups, i.e.

1. *Government institutions*, which handle fundings, governances, and regulations;
2. *Academic institutions*, which handle research, innovation, and education;
3. *International organizations*, which engage stakeholders from public and private sectors, both locally and globally;
4. *Industries*, which may develop or use quantum technologies;
5. *Individuals* who may develop or consume quantum technologies.

This is very similar to the quadruple helix model (Carayannis & Campbell, 2009) used by Gercek and Seskir (2025), consisting of government, academia, industry, and civil society. In our opinion, identifying and understanding the relationships between the stakeholders within the regional quantum ecosystem is a very important first step of engagement. We adopt the finer classification of stakeholders from WEF, mainly because knowledge on quantum science and technologies is not readily available for non-STEM individuals in the region. We recognize that more efforts in quantum education and science communication are needed in Section 4.4.

According to Vermaas (2025), "quantum" is often regarded as *unintelligible*. In order to understand quantum technologies, some knowledge on quantum physics is usually required. However, this high entry barrier has impeded conversations around it. As a mean to facilitate the discussion, several projects were being carried out, such as the Not Art, Not Quantum workshop[34] and Quantum Art Thailand[35], intersecting quantum science, art, and gamification (Choong et al., 2025). Particularly, we are curious on how quantum can emerge in art and gamification, or how art and gamification assist the general public to understand quantum physics and its nuances without using analogies, which are often inaccurate at the fundamental level. Figure 3 shows an example of our future work on how art may enhance the interpretation of quantum physics and vice versa. Of course, there are limitations on using art and gamification to describe quantum physics as well, which will be studied closely (Choong & Nik Aimi, forthcoming).

When engaging stakeholders, a quote from Richard Feynmann "*I think I can safely say that nobody understands quantum mechanics*" is often taken out of context to highlight the unintelligibility of quantum physics. It should be emphasized that the original context of the quote was meant to describe the *non-intuitiveness* of quantum physics. The mathematical formulation of quantum physics produces the right results and is very well understood by many physicists. Providing the correct contexts and interpretation of quantum physics will build public trust in quantum technologies.

---

[34] https://open-quantum-institute.cern/not-art-not-quantum-bridging-creativity-with-quantum/
[35] https://www.facebook.com/QuantumArtTH/

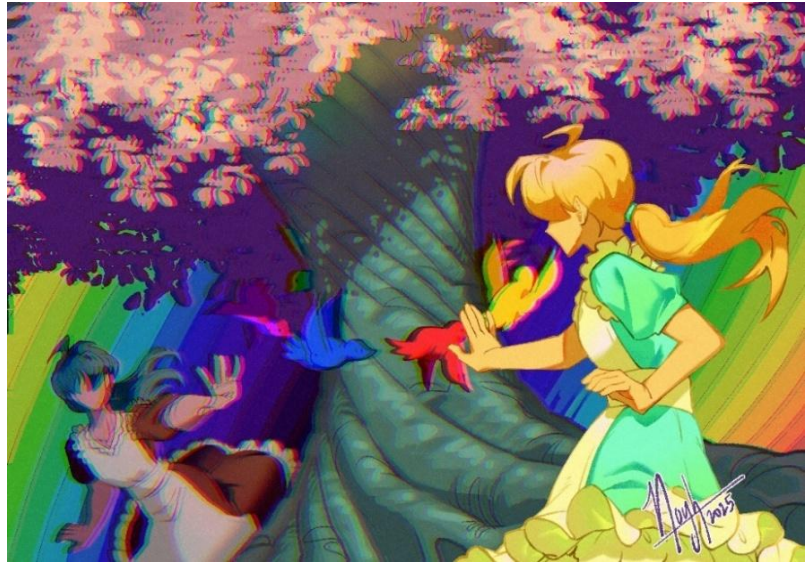

Figure 3: *Lost in Fibration*, an artwork co-created by Nik Aimi Afiqa and Pak Shen Choong. From the quantum-art intersection, it depicts a storytelling of *Alice in Wonderland*, making use of colour theory and visual effects to illustrate the Hopf fibration of one qubit, Pauli basis, and quantum superposition[36].

**4.3 Establishing policies, benchmarks, and coordinating committee**

Collingridge's dilemma (1980) describes the contradicting dynamics of innovation and control. During the early stage of innovation, it is easy to apply control and regulation, albeit the impacts of said technology may not be easy to predict. At this stage, an over-restricting regulation may interfere and slow down the innovation process. Whereas, when the technology becomes widespread, its impacts will be apparent, though implementing regulations will also become more challenging. Artificial intelligence sets up a good example where innovation outpaces regulation (Feng, Cramer, & Mans, 2022).

Standing at this crossroad, it is vital to reinforce the implementation of responsible research and innovation (RRI) (Schmidt et al., 2025) into the national and regional quantum roadmap. Together with the UNESCO Recommendation on Open Science (2021), the ethical assessment framework provides a transparent and responsible development of quantum technologies that minimizes the downsides of socio-economical disruptions. One thing we would like to take note is that under the UNESCO's recommendation, *open science* does not mean that scientific knowledge and resources are shared without any limits, but rather *as open as possible*. National security, IP protections, and data ethics may restrict access to certain information and resources. Another consideration is to make use of anticipation and foresight to identify potential opportunities, challenges, and risks. Policy making and regulatory practices have to be proactive and flexible, so that they reflect to the current issues and discussions in the global quantum landscape (GESDA, 2025).

---

[36] https://www.linkedin.com/pulse/lost-fibration-pak-shen-choong-fzt8f/

For quantum computers, their energy landscape and utility-scale are not well-defined. As stated before, the quantum advantage for noisy intermediate-scale quantum (NISQ) algorithms is not fully explored and systematically studied (Huang et al., 2025). Several benchmarking initiatives and metric to provide a standardization on quantum technologies were proposed, for instance the Quantum Energy Initiative (Auffèves, 2022), quantum volume (Cross et al., 2019), and Quantum Benchmark Initiative[37]. A regional coordinating committee may be established to monitor and benchmark the regional quantum development and prevent low-quality quantum research.

**4.4 Democratizing quantum education**

Quantum research is a nascent and fast-moving field. Despite that quantum physics was developed in a century ago, new discoveries on quantum technologies, quantum algorithms, and industrial use cases are constantly being made. On the other hand, the accreditation of a quantum information course for Bachelor degree will take an average of three to five years before the program is available. It is important to democratize quantum education to improve quantum literacy for everyone. This has been done through non-profit efforts, such as QWorld[38], the education consortium of OQI, and a collaborative effort [39] by the International Telecommunication Union (ITU), Quantum Delta NL, and United Nations International Computing Centre (UNICC).

Democratizing quantum education also means providing access to quantum knowledge for non-physics students. Quantum use case hackathons, such as SEA Quantathon 2025 and QAI GenQ Hackathon 2025 provide opportunities for students to develop an industrial application of quantum algorithms, complemented by a business and SDGs pitch. This allows students from different disciplinaries to participate and co-create a quantum use case for the relevant industries, and break the silos that quantum research is only for physics students.

**4.5 Creating multiple career pathways for quantum graduates**

As stated by OECD (2025), most of the available quantum jobs require a PhD degree. Meanwhile, quantum technologies have the potentials to penetrate multiple industries, such as communication, logistics, finance, metrology, cybersecurity, healthcare, renewable energy, and artificial intelligence. Most of these applications do not require a PhD in quantum physics but rather a multi-disciplinary approach to understand the societal, technological, industrial, and environmental impacts of quantum technologies. This reveals that there is a need for quantum scientists to work outside of their expertise to promote the industrial adoption of quantum technologies and encourage entrepreneurship through quantum use case development.

---

[37] https://www.darpa.mil/research/programs/quantum-benchmarking-initiative
[38] https://qworld.net/
[39] https://www.quantum-course.com/

According to the Malaysian Graduate Tracer Study 2025[40], as of April 2025, 87.5% of all graduates had secured employment, and 35% of those hired were working in jobs unrelated to their field of study. To put into a better perspective, 38.75% of graduates were either working in jobs unrelated to their field of study or simply unemployed. While there is a lack of systematic study to support this claim, brain drain has been a serious problem in Southeast Asia[41]. The issues of skill misalignment and unemployment rate must be tackled promptly by multiple approaches, such as enabling the quantum supply chain, i.e. cryogenics, qubit control electronics, photonics, and semiconducting chips, or establishing venture capitals to support quantum startups as an early form of a regional quantum industry.

## 5.0 Conclusion

As a form of tech diplomacy, quantum diplomacy evolves from science diplomacy out of the necessity to ensure the responsible development and deployment of quantum technologies, for the benefits of humanity. Quantum diplomacy engages various stakeholders in neutral dialogues to resolve hard power problems and bridge the quantum divide, which is vital for a newly established regional quantum ecosystem. In this paper, we proposed four recommendations to strengthen this regional community, with the goals of

1. mitigating quantum hype and misinformation;
2. establishing standards and benchmarking metric
3. ensuring equitable access to quantum education;
4. assisting regional market penetration.

We note that our recommendations are derived from the perspective of scientists, for scientists. As such, this paper does not consider the roles and opportunities from regional state actor such as ASEAN, and international or intergovernmental organizations such as United Nations Economic and Social Commission for Asia and the Pacific (UN ESCAP), International Science, Technology and Innovation Center (ISTIC), Asia-Pacific Economic Cooperation (APEC), or Economic Research Institute for ASEAN and East Asia (ERIA). Since most of the Southeast Asia countries are in the process of developing a national quantum strategy, a systematic comparison and analysis of the quantum roadmap development between different Southeast Asia countries will be left as a future work.

---

[40] https://graduan.mohe.gov.my/skpg25/
[41] https://fulcrum.sg/aseanfocus/asean-can-help-to-address-brain-drain-in-southeast-asia/

**Data availability statement:** Not applicable.

**Financial interest:** Pak Shen Choong has received a full scholarship to join the Geneva Science Diplomacy Week 2025.

**Funding:** Nurisya Mohd Shah would like to acknowledge Universiti Putra Malaysia Grant GPI/2024/9800400.

**Authors' contributions:** Pak Shen Choong drafted the work and made substantial contributions to the conception or design of the work. Nurisya Mohd Shah revised the work critically for important intellectual content. Yung Szen Yap acquired and analysed Malaysia's quantum landscape critically.

**Acknowledgements**

Pak Shen Choong would like to thank Geneva Science and Diplomacy Anticipator (GESDA) and Open Quantum Institute (OQI) for their supports in realizing the Southeast Asia quantum ecosystem. Pak Shen Choong would like to thank Clarissa Ai Ling Lee for insightful discussions on quantum artscience, Zeki Seskir, Clara Yun Fontaine, Angelina Frank, Marga Gual Soler, and Marianne T. Schörling for the illuminating discussions on responsible quantum technologies, quantum governance, and quantum diplomacy. The authors acknowledge the regional efforts provided by Worawat Meevasana and Areeya Chantasri from Thailand, Bobby Corpus, Dylan Josh Lopez, and Jabez Ayson from the Philippines, Nguyen Quoc Hung and Rossy Nhung Nguyen from Vietnam, Whei Yeap Suen and Leong Chuan Kwek from Singapore, and lastly Khoirul Anwar and Yanoar Pribadi Sarwono from Indonesia.